\documentclass[12pt]{article}
\usepackage[dvips]{graphicx}
\usepackage{amsfonts,amsthm,amstext,amscd}
\usepackage{amssymb}
\usepackage{yfonts}

\begin{document}
\baselineskip=15.5pt
\pagestyle{plain}
\setcounter{page}{1}

\par\hfill KUL-TF-10/06

\vspace{2cm}

\begin{center}
{\LARGE{\bf An elementary stringy estimate of transport coefficients of large temperature QCD}}
\end{center}
\vskip 10pt
\begin{center}
{\large 
Francesco Bigazzi and Aldo L. Cotrone
}
\end{center}
\vskip 10pt
\begin{center}
\textit{Institute for theoretical physics, K.U. Leuven;
Celestijnenlaan 200D, B-3001 Leuven,
Belgium.}\\

{\small francesco.bigazzi@fys.kuleuven.be, aldo.cotrone@fys.kuleuven.be}
\end{center}

\vspace{15pt}

\begin{center}
\textbf{Abstract}
\end{center}

\vspace{4pt}{\small \noindent 
Modeling QCD at large temperature with a simple holographic five dimensional theory encoding minimal breaking of conformality, allows for the calculation of all the transport coefficients, up to second order, in terms of a single parameter.
In particular, the shear and bulk relaxation times are provided.
The result follows by deforming the AdS background with a scalar dual to a marginally relevant operator, at leading order in the deformation parameter.
}
\vfill

\newpage
\section{Introduction and results}

The evolution of the Quark-Gluon Plasma (QGP) produced in heavy ion collisions is fairly well described, after a short period of thermalization and before hadronization, by relativistic hydrodynamics, consistently with the strongly coupled regime of such system \cite{arsene}.
Numerical simulations of the hydrodynamic evolution of the QGP require as input the value of the transport coefficients.
While recent simulations indicate that the evolution of the QGP should be quite insensitive to most of the second order coefficients (see for example \cite{Luzum:2008cw,Rajagopal:2009yw,Song:2009rh}), it is definitely sensitive to the value of the shear viscosity \cite{lista}, and can be influenced in a sizable way by the bulk viscosity and possibly the relaxation times \cite{Rajagopal:2009yw,Song:2009rh,Denicol:2010tr}.
Moreover, a complete characterization of the Quark-Gluon Plasma of QCD up to second order still requires the knowledge of the whole set of coefficients.

There are currently no first-principle reliable calculations of almost all the second order coefficients for QCD at strong coupling: lattice results give some estimates of the viscosities and the shear relaxation time \cite{Meyer:2007dy}, but they are affected by considerable uncertainties (see for example \cite{Moore:2008ws}).
In fact, actual simulations, lacking solid data for QCD, make often use, as benchmark values of the transport coefficients, of the ones derived from the gravitational dual of ${\cal N}=4$ SYM \cite{Baier:2007ix} (in some cases together with the bound on the bulk viscosity proposed in \cite{buchelbound} and a relation for the relaxation times from \cite{Romatschke:2009kr}). 
While the ${\cal N}=4$ SYM values for the ``shear'' coefficients are expected to be in the right ballpark for QCD, they still concern an exactly conformal theory, and in particular the bulk viscosity and many of the second order coefficients are not determined.

In order to improve this situation, the first step is to break conformal invariance. Since QCD is approximately conformal in the temperature window $1.5 T_c \lesssim T \lesssim 4 T_c$, the conformality breaking effects can be treated perturbatively.
In this situation, probably the simplest way of modeling QCD holographically is by a theory where conformality is slightly broken by a marginally relevant operator.

The aim of this note is to point out that, in such a scenario, all the transport coefficients up to second order for the uncharged plasma are given in terms of a single parameter (weighting the conformality breaking) by making use of the results in \cite{Kanitscheider:2009as,Romatschke:2009kr}: they are collected in Table \ref{relations}.
In particular, the behavior of the shear and bulk relaxation times is briefly discussed in section \ref{secresults}.

There are surely more precise ways of modeling holographically QCD (none of which is of course completely correct). 
Nevertheless, the model considered in this note has the considerable advantage of full calculability, providing one of the few examples in which all the second order transport coefficients are determined.
Moreover, the results in Table \ref{relations} hold for any theory with gravity dual, where conformality is broken at leading order by a marginally (ir)relevant operator (dual to a scalar with the simplest possible potential (\ref{potential})), including the cascading plasma \cite{Gubser:2001ri} and the D3D7 plasmas \cite{Bigazzi:2009bk}. 

\subsection{Notation}
Uncharged relativistic hydrodynamics is determined, up to second order in the derivative expansion, by seventeen transport coefficients, fifteen of which are possibly independent \cite{Baier:2007ix}, \cite{Romatschke:2009kr}.
On a general space with metric $g_{\mu\nu}$, the energy momentum tensor
\begin{equation}\label{tmunu}
T^{\mu\nu}=\varepsilon u^\mu u^\nu + p  \Delta^{\mu\nu} + \pi^{\mu\nu} + \Delta^{\mu\nu}\Pi \,, \qquad {\rm where}\qquad  \Delta^{\mu\nu}=g^{\mu\nu}+u^\mu u^\nu\,,
\end{equation}
is determined by the energy density $\varepsilon$, fluid velocity $u^\mu$ ($u^\mu u_\mu=-1$), the transport coefficients in its ``viscous shear'' part:
\begin{eqnarray}\label{shear}
\pi^{\mu\nu}&=&-\eta \sigma^{\mu\nu} +\eta \tau_\pi \Bigl[\langle D \sigma^{\mu\nu}\rangle + \frac{\nabla \cdot u}{3}\sigma^{\mu\nu} \Bigr] + \kappa \Bigl[ R^{<\mu\nu>}-2 u_\alpha u_\beta R^{\alpha <\mu\nu> \beta} \Bigr] + \lambda_1 \sigma^{<\mu}_{\lambda} \sigma^{\nu>\lambda}  \nonumber \\
&&+ \lambda_2 \sigma^{<\mu}_{\lambda} \Omega^{\nu>\lambda} + \lambda_3 \Omega^{<\mu}_{\quad \lambda} \Omega^{\nu>\lambda}  + \kappa^* 2 u_\alpha u_\beta R^{\alpha <\mu\nu> \beta}   \nonumber \\
&& + \eta \tau_\pi^* \frac{\nabla \cdot u}{3}\sigma^{\mu\nu} + \lambda_4 \nabla^{<\mu} \log{s}  \nabla^{\nu >} \log{s}
\end{eqnarray}
and in its ``viscous bulk'' part:
\begin{eqnarray}\label{bulk}
\Pi &=&-\zeta (\nabla \cdot u) + \zeta \tau_\Pi D  (\nabla \cdot u) + \xi_1  \sigma^{\mu\nu}\sigma_{\mu\nu}+ \xi_2  (\nabla \cdot u)^2 + \xi_3 \Omega^{\mu\nu}\Omega_{\mu\nu} + \xi_4 \nabla_{\mu}^{\perp} \log{s} \nabla^{\mu}_{\perp} \log{s} \nonumber \\
&& + \xi_5 R + \xi_6 u^\alpha u^\beta R_{\alpha \beta}\,,
\end{eqnarray}
while the pressure is given by the equation of state $p(\varepsilon)$.
The various structures in (\ref{shear}) and (\ref{bulk}), apart from the obvious Riemann and Ricci tensors and scalar curvature ($R^{\mu\nu\rho\sigma}, R^{\mu\nu}, R$), are given by:
\begin{eqnarray}
D &\equiv & u^\mu\nabla_\mu\,, \qquad \nabla^{\mu}_{\perp} \equiv  \Delta^{\mu\nu}  \nabla_{\nu}\,, \qquad \sigma^{\mu\nu}\equiv  \nabla^{\mu}_{\perp} u^\nu + \nabla^{\nu}_{\perp} u^\mu -\frac23  \Delta^{\mu\nu}(\nabla \cdot u)\,, \nonumber\\ 
\Omega^{\mu\nu} &\equiv & \frac12 (\nabla^{\mu}_{\perp} u^\nu - \nabla^{\nu}_{\perp} u^\mu)\,, \qquad 
\end{eqnarray}
and for a generic tensor $A^{\mu\nu}$ it was used the notation:
\begin{equation}
\langle A^{\mu\nu} \rangle = A^{<\mu\nu>} \equiv \frac12  \Delta^{\mu\alpha}\Delta^{\nu\beta}(A_{\alpha\beta}+A_{\beta\alpha})-\frac13  \Delta^{\mu\nu} \Delta^{\alpha\beta}A_{\alpha\beta}\,.
\end{equation}
Finally, $s$ is the entropy density, while the speed of sound will be denoted as $c_s^2=dp/d\varepsilon$.

The shear viscosity $\eta$ and the second order coefficients $\tau_\pi$ (``shear'' relaxation time), $\kappa$, $\lambda_1, \lambda_2, \lambda_3$ are the only ones defined in conformal fluids, as the one of ${\cal N}=4$ SYM.
All the others coefficients, i.e. the bulk viscosity $\zeta$ and the second order coefficients $\kappa^*, \tau_\pi^*, \lambda_4, \tau_\Pi$ (``bulk'' relaxation time), $\xi_1,   \xi_2,  \xi_3,  \xi_4,  \xi_5,  \xi_6$, are only defined in non-conformal plasmas.

\subsection{The estimate}\label{secresults}

\setcounter{table}{0}

Consider a gravity dual model for QCD at large temperature, where the leading conformality breaking effect is captured by adding to the five dimensional metric a non trivial dilaton profile, dual to a marginally relevant operator.
The main observation of this note is that, for the simplest scalar potential, the transport coefficients are completely determined in terms of a single parameter.
Defining:
\begin{equation}\label{Delta}
\delta \equiv (1-3c_s^2)\,, 
\end{equation}
at first order in $\delta$ the transport coefficients are given in Table \ref{relations}.
\begin{table}[h]
\begin{center}
\begin{tabular}{||c|c||c|c||c|c||} 
\hline
 & & & & & \\
$ \frac{\eta}{s} $ & $\frac{1}{4\pi}$ &  $T\tau_{\pi}  $  & $ \frac{2-\log{2}}{2\pi} + \frac{3(16-\pi^2)}{64\pi}\delta $  & $ \frac{T\kappa}{s} $  &  $  \frac{1}{4\pi^2}\Bigl(1-\frac34 \delta \Bigr) $  \\
 & & & & & \\
\hline \hline
 & & & & & \\  
$\frac{T \lambda_1}{s}  $ & $\frac{1}{8\pi^2}\Bigl(1+\frac34 \delta \Bigr) $ & $\frac{T \lambda_2}{s} $ & $-\frac{1}{4\pi^2}\Bigl( \log{2}+\frac{3\pi^2}{32}\delta \Bigr) $ & $\frac{T \lambda_3}{s} $ & $0 $ \\ 
 & & & & & \\
\hline \hline
 & & & & & \\
$\frac{T\kappa^*}{s} $ & $-\frac{3}{8\pi^2}\delta $ & $T\tau_{\pi}^* $ & $-\frac{2-\log{2}}{2\pi}\delta $ & $\frac{T \lambda_4}{s}  $ & $0 $ \\
 & & & & & \\
\hline \hline
 & & & & & \\
$\frac{\zeta}{\eta} $ & $\frac23 \delta $ & $T\tau_{\Pi} $ & $\frac{2-\log{2}}{2\pi} $ & $\frac{T \xi_{1}}{s} $ & $\frac{1}{24\pi^2}\delta $ \\
 & & & & & \\
\hline \hline
 & & & & & \\
$ \frac{T \xi_{2}}{s} $ & $\frac{2-\log{2}}{36\pi^2}\delta $ & $\frac{T \xi_{3}}{s} $ & $0 $ & $\frac{T \xi_{4}}{s} $ & $0 $ \\
 & & & & & \\
\hline \hline
 & & & & & \\
$\frac{T \xi_{5}}{s} $ & $\frac{1}{12\pi^2}\delta $ & $\frac{T \xi_{6}}{s} $ & $\frac{1}{4\pi^2}\delta $ & & \\
 & & & & & \\
\hline
\end{tabular} 
\end{center}
\caption{The transport coefficients, in the notation of (\ref{tmunu})-(\ref{bulk}), for a marginally (ir)relevant deformation of a conformal theory, at leading order in the deformation parameter $\delta \equiv (1-3c_s^2)$. The holographic equation of state is $\varepsilon=3(1+\delta)p$.}\label{relations}
\end{table}
This result, which is the main content of this note, follows directly from \cite{Kanitscheider:2009as}, \cite{Romatschke:2009kr} (which already contains a part of the relations in Table \ref{relations}\footnote{See also \cite{tutti1,buchelbound,gubserspeed,tutti2,Kanitscheider:2009as,tutti3,gubserspeed2,noi}.}) and will be derived in section \ref{proof}.

Possibly the main novel results contained in Table \ref{relations} concern the two relaxation times $\tau_{\pi}, \tau_{\Pi}$.
Specifically, at leading order in the conformality breaking, the bulk relaxation time $\tau_{\Pi}$ is \emph{not} proportional to the bulk viscosity.
The behavior of the shear relaxation time $\tau_{\pi}$ is instead more interesting, since it depends on the speed of sound.
For a phenomenologically realistic behavior of the latter, $\tau_{\pi}$ is \emph{decidedly increasing} when reducing the temperature.
In particular, it increases \emph{faster} than $\tau_{\Pi}$. 

Moreover, using the above results it is easy to verify that the relation
\begin{equation}
4\,\lambda_1 + \lambda_2 = 2\,\eta\, \tau_{\pi}\,,
\label{relhy}
\end{equation}
holds, at first order in $\delta$. It has been shown in \cite{erd,hy} that (\ref{relhy}) is satisfied in all the known examples of conformal plasmas (in $d\ge 4$ spacetime dimensions, with of without conserved global charges) with dual gravity description. Our results provide a unique validity check of the above relation in non-conformal settings.\footnote{We thank Todd Springer for this observation.} 

While the present system can model at best the regime of QCD away from the critical temperature, where there are certainly other ways of modeling QCD, it would be unexpected if the qualitative behavior of the transport coefficients described above turned out to be drastically different. 

In order to give an illustrative example of numerical estimates of the coefficients, we have to chose one input parameter.
As in \cite{Song:2009rh}, we use the results for the speed of sound from the lattice study in \cite{Katz:2005br}.
We consider a temperature $T\sim 1.5 T_c$, which is a reasonable value for the RHIC experiment.
Then from \cite{Katz:2005br} we read $c_s^2\sim 0.283$ from which we get the numbers in Table \ref{heiz1}.
\begin{table}[h] 
\begin{center}
\begin{tabular}{||c|c||c|c||c|c||}
\hline
 & & & & & \\
$ \frac{\eta}{s} $ & $\frac{1}{4\pi}$ &  $T\tau_{\pi}  $  & $0.222 $  & $ \frac{T\kappa}{s} $  &  $0.022 $  \\
 & & & & & \\
\hline \hline
 & & & & & \\  
$\frac{T \lambda_1}{s}  $ & $0.014 $ & $\frac{T \lambda_2}{s} $ & $-0.021 $ & $\frac{T \lambda_3}{s} $ & $0 $ \\ 
 & & & & & \\
\hline \hline
 & & & & & \\
$\frac{T\kappa^*}{s} $ & $-0.006 $ & $T\tau_{\pi}^* $ & $-0.031 $ & $\frac{T \lambda_4}{s}  $ & $0 $ \\
 & & & & & \\
\hline \hline
 & & & & & \\
$\frac{\zeta}{\eta} $ & $0.101 $ & $T\tau_{\Pi} $ & $0.208 $ & $\frac{T \xi_{1}}{s} $ & $0.001 $ \\
 & & & & & \\
\hline \hline
 & & & & & \\
$ \frac{T \xi_{2}}{s} $ & $0.001 $ & $\frac{T \xi_{3}}{s} $ & $0 $ & $\frac{T \xi_{4}}{s} $ & $0 $ \\
 & & & & & \\
\hline \hline
 & & & & & \\
$\frac{T \xi_{5}}{s} $ & $0.001 $ & $\frac{T \xi_{6}}{s} $ & $0.004 $ & & \\
 & & & & & \\
\hline
\end{tabular} 
\end{center}
\caption{The transport coefficients at $T\sim 1.5 T_c$ and $c_s^2 \sim 0.283$.} \label{heiz1}
\end{table}
The reported values provide corrections up to $17\%$ to the conformal ones (when the latter are defined).
In particular $2\pi T\tau_\pi= 1.394$ is a bit larger than the conformal value (1.307) and more similar to the one used in \cite{Song:2009rh}. The numerical difference is by definition not very large, but sizable.
Obviously, increasing the temperature reduces this difference and at $T \sim 3 T_c$, which could be a significant temperature for LHC, the corrections to the conformal values are below $10\%$.

Let us conclude this section by describing the approximations involved in applying these relations to QCD. 
First of all, QCD does not have a purely gravitational dual.
Nevertheless, experience teaches that simple gravity models of metric plus scalar are in good quantitative agreement with the lattice results for certain observables.
In particular, we are interested in the regime, relevant in the early stages of the QGP evolution, at $T>T_c$ where QCD is nearly conformal and strongly coupled.
Moreover, in the hydrodynamic regime the gravity description and actual QCD are in good agreement (e.g. the result for the shear viscosity).
On the other hand, in QCD the gluon condensate is marginally relevant in the asymptotically free regime, while in the experimental regime we are interested in, the theory is strongly coupled and this operator can be expected to have developed a sizable anomalous dimension.\footnote{This situation could be modeled with a scalar dual to a relevant operator \cite{gubserspeed,gubserspeed2}. In this case the computation of the second order coefficients is highly more complicated.}
The other caveat concerns the effects of the flavors and the chemical potential, which are not accounted for in Table \ref{relations}, but are expected to give the latter subleading corrections.

In view of these considerations, the relations in Table \ref{relations} can provide a fair estimate\footnote{Better than the one provided by ${\cal N}=4$ SYM \cite{Baier:2007ix}.} of the initial behavior of the hydrodynamic evolution at RHIC and LHC.

\section{Derivation}\label{proof}

The leading conformality breaking effects of a source for a marginally (ir)relevant operator can be captured in the dual gravitational setting by a so-called Chamblin-Reall model.
Consider an effective five dimensional theory with metric plus a single scalar $\phi$ with potential $V(\phi)$.
It models the breaking of conformality at leading order in a small parameter $\epsilon$ if $V(\phi)|_{\epsilon=0}=V_0$, where the negative cosmological constant $V_0$ allows for and $AdS$ solution. 
The operator dual to $\phi$ (on the unperturbed $AdS$ solution) is of dimension four, that is it is marginally (ir)relevant, if $\partial_{\phi}^2 V(\phi)|_{\phi=0}={\cal O}(\epsilon^{1+\alpha})$ with positive $\alpha$.
The simplest such class of models, and the one we are interested in, is given by:
\begin{equation}\label{potential}
V(\phi) = V_0 + \epsilon \phi + {\cal O}(\epsilon^{1+\alpha})\,.
\end{equation}
At leading order:
\begin{equation}
V(\phi) \sim V_0 e^{\epsilon \phi/V_0}\,,
\end{equation}
i.e. the model is in the Chamblin-Reall class \cite{Chamblin:1999ya}.\footnote{To be precise, with unit $AdS$ radius, $V_0=-12$ and, from the calculation of the speed of sound, $\epsilon^2=96\delta$ \cite{gubserspeed}.}

For this class of models, the proof of the relations in Table \ref{relations} follows directly from \cite{Kanitscheider:2009as}.
Let us summarize it.
The starting point is the fact that Chamblin-Reall models in $d+1$ dimensions, for particular values of the coefficient of the exponential in the potential, can be obtained from dimensional reduction on a $2\sigma-d$ torus of pure gravity plus cosmological constant in $2\sigma+1$ dimensions.
This happens when the parameter $\sigma$, which determines together with $d$ the coefficient in the exponential, is semi-integer.
For these values of $\sigma$, one can then start from the well-known $AdS_{2\sigma+1}$ solution and its dual hydrodynamic energy-momentum tensor, and obtain the hydrodynamic energy-momentum tensor for the dual to the Chamblin-Reall model by simple toroidal dimensional reduction.

The crucial observation in \cite{Kanitscheider:2009as} is that, from the point of view of the theory in $d+1$ dimensions, the equations are smooth in the parameter $\sigma$.
This allows for the computation of the hydrodynamic energy-momentum tensor for arbitrary values of $\sigma>d/2$.\footnote{At $\sigma=d/2$ the action is singular \cite{Kanitscheider:2009as}.}

The procedure is as follows.
One starts from a Chamblin-Reall model in $d+1$ dimensions for whatever $\sigma>d/2$ and performs the continuation (which is smooth) to the nearest value $\tilde\sigma$ which is semi-integer.
The latter theory is the compactification of a theory admitting a $AdS_{2\tilde\sigma+1}$ solution, so its dual energy-momentum tensor, which will be a function of $\tilde\sigma$, can be calculated straightforwardly.
This  energy-momentum tensor can thus be continued (smoothly) back to the one of the theory corresponding to the original value $\sigma$.

In particular, all the transport coefficients for this theory will automatically be determined by the conformal ones in the higher dimensional theory, modulo an overall constant (the volume of the torus) which can be fixed knowing just one coefficient.

Let us see concretely how this procedure is implemented. 
One can determine $\sigma$ by the fact that the equation of state in these models is $\varepsilon=(2\sigma-1)p$ \cite{Kanitscheider:2009as}, so that $\sigma=2+3\delta/2$ in the notation of Table \ref{relations}.\footnote{And $\sigma=2+\epsilon^2/64$ in the notation of (\ref{potential}).}
Thus, for a small deformation $\delta$ of a conformal theory, $\tilde\sigma=5/2$ and the relevant starting solution is $AdS_{2\tilde\sigma+1}=AdS_{6}$, whose dual conformal hydrodynamics was considered in \cite{Bhattacharyya:2008mz}.
Let us write the results of  \cite{Bhattacharyya:2008mz} in the present notation:
\begin{eqnarray}\label{ads6}
\eta^{(2\tilde\sigma)}&=&\frac{s^{(2\tilde\sigma)}}{4\pi}\,, \nonumber \\ 
\kappa^{(2\tilde\sigma)}&=&\frac{\eta^{(2\tilde\sigma)}\tilde\sigma}{\pi T(2\tilde\sigma-2)}\,, \nonumber \\ 
\tau_{\pi}^{(2\tilde\sigma)}&=&\frac{\tilde\sigma}{2\pi T}\Bigl(1-\int_1^{\infty} \frac{y^{2\tilde\sigma-2}-1}{y(y^{2\tilde\sigma}-1)}dy\Bigr) \,,\nonumber \\ 
\lambda_1^{(2\tilde\sigma)}&=& \frac{\eta^{(2\tilde\sigma)}\tilde\sigma}{4\pi T}\,, \nonumber \\ 
\lambda_2^{(2\tilde\sigma)}&=& -\frac{\eta^{(2\tilde\sigma)}\tilde\sigma}{\pi T}\int_1^{\infty} \frac{y^{2\tilde\sigma-2}-1}{y(y^{2\tilde\sigma}-1)}dy \,, \nonumber \\ 
\lambda_3^{(2\tilde\sigma)}&=& 0\,.
\end{eqnarray}
The procedure to obtain the desired coefficients involves reducing the energy momentum tensor on a circle ($2\tilde\sigma-d=1$) of volume $V$, continuing it back to $\sigma$ and expanding it at first order in $\delta$ \cite{Kanitscheider:2009as}; examples of results of this procedure are (the arrows denote the analytic continuation):
 \begin{eqnarray}\label{reduction}
\eta &=&\eta^{(2\tilde\sigma)}V \quad \rightarrow \quad \eta^{(2\tilde\sigma)}V \,, \\ 
\kappa &=&\kappa^{(2\tilde\sigma)}V\quad \rightarrow \quad \frac{\eta (4+3\delta)}{2\pi T (2+3\delta)} \sim  \frac{\eta}{\pi T}\Bigl( 1-\frac{3}{4}\delta \Bigr) \,,  \\ 
\tau_{\pi} &=&\tau_{\pi}^{(2\tilde\sigma)}V \quad \rightarrow \quad \frac{(4+3\delta)V}{4\pi T}\Bigl(1-\int_1^{\infty} \frac{y^{2+3\delta}-1}{y(y^{4+3\delta}-1)}dy\Bigr)\sim \frac{V}{\pi T}\Bigl[ \frac{2-\log{2}}{2}+\frac{3(16-\pi^2)}{64}\delta \Bigr] \,,\nonumber\label{taupi} \\ \\ 
\lambda_1 &=& \lambda_1^{(2\tilde\sigma)}V \quad \rightarrow \quad \frac{\eta}{2\pi T}\Bigl( 1+\frac{3}{4}\delta \Bigr) \,, \\ 
\lambda_2 &=& \lambda_2^{(2\tilde\sigma)}V \quad \rightarrow \quad  -\frac{\eta (4+3\delta)}{2\pi T} \int_1^{\infty} \frac{y^{2+3\delta}-1}{y(y^{4+3\delta}-1)}dy \sim -\frac{\eta}{\pi T}\Bigl( \log{2}+\frac{3\pi^2}{32}\delta \Bigr) \,, \\ 
\lambda_3 &=& 0\,,\\
\zeta & = & 2\eta^{(2\tilde\sigma)}V\frac{2\tilde\sigma-d}{(2\tilde\sigma-1)(d-1)} \quad \rightarrow \quad 2\eta\frac{3\delta}{3(3+3\delta)}\sim \frac{2}{3}\eta \delta\,,
\end{eqnarray}
where the leading ``conformal'' term in (\ref{taupi}) fixes the value $V=1$ \cite{Baier:2007ix}.
The other coefficients in Table \ref{relations} are obtained in the same way.

Let us conclude by stressing again the fact that the relations in Table \ref{relations} are valid for any theory where conformality is broken at leading order by a marginally (ir)relevant deformation, with the dual scalar having the potential (\ref{potential}).
These theories\footnote{For a recent study of their shear spectral sum rule see \cite{spr}.} include the cascading plasmas \cite{Gubser:2001ri} and the D3D7 plasmas \cite{Bigazzi:2009bk}. 
The latter are the first examples of holographic plasmas including the effects of dynamical flavors in a completely controllable framework.
In this case, the relations in Table \ref{relations} match precisely the coefficients calculated in \cite{noi}\footnote{The small parameter in \cite{noi} was denoted as $\epsilon_h^2=6\delta$.} and complete the determination of all the second order transport coefficients in those plasmas.

\vskip 15pt
\centerline{\bf Acknowledgments}
\vskip 10pt
\noindent
We are grateful to D. Mayerson, T. Springer and J. Tarrio for discussions. This work is supported by the FWO -Vlaanderen, project G.0235.05 and by the Federal Office for Scientific, Technical and Cultural Affairs through the Interuniversity Attraction Poles Programme (Belgian Science Policy) P6/11-P.

{ \it F. B. and A. L. C. would like to thank the Italian students, parents, teachers and scientists for
their activity in support of public education and research.}

\end{document}